\documentstyle[aps,prl]{revtex}

\begin{document}
\draft

\vspace{1cm}

\title{Comment on ``Theorem for nonrotating singularity-free universes''}

\maketitle

A.K. Raychaudhuri presented recently\cite{lett} a general theorem 
stating that ``in any singularity-free nonrotating universe, open in
all directions, the spacetime average of all stress energy
invariants including the energy density vanish''. The strong energy
condition is assumed. The proof was obtained by taking the spacetime
average of both sides of the Raychaudhuri equation
\begin{equation}
\label{eq}
\dot{\theta} - \nabla_a\dot{\xi}^a = - \sigma_{ab}\sigma^{ab} -
\frac{1}{3}\theta^2 + \omega_{ab}\omega^{ab} - R_{ab}\xi^a\xi^b,
\end{equation}
where $\xi^a$ is the unit time-like vector along the world lines of
matter. For nonrotating universes $\omega^{ab}=0$, and with the
assumption of the strong energy condition each term in the R.H.S. of
(\ref{eq}) is non-positive. By using general arguments,  he could
 show
that $\langle \dot{\theta}\rangle =\langle \nabla_a\dot{\xi}^a\rangle = 0$,
implying that the spacetime average of each term in the R.H.S. of
(\ref{eq}) vanishes. According to the Letter, this result suggests
that one should  give up of a realistic singularity-free description
of the universe.

We stress here that, even 
for realistic models with a initial singularity,
the spacetime average of the invariants in the R.H.S. of (\ref{eq})
can vanish, and hence, this cannot be used to rule out singularity-free
models. Let us take, for instance, the Robertson-Walker line element
with $\kappa =0$, $ds^2 = -dt^2  + a^2(t) \left(dx^2 + dy^2 + dz^2 \right)$.
With $\xi^a = (1,0,0,0)$, we have $\dot{\xi}^a = 0$ identically and
\begin{eqnarray}
\langle \dot{\theta}\rangle &=& 
\lim_{\stackrel{t_1,x_1,y_1,z_1\rightarrow\infty}{t_0\rightarrow 0}}
\frac{\int_{t_0}^{t_1} \int_{-x_1}^{x_1} \int_{-y_1}^{y_1} \int_{-z_1}^{z_1} 
\left(\xi^a \nabla_a \theta \right)\sqrt{g}\, d^4 x}
{\int_{t_0}^{t_1} \int_{-x_1}^{x_1} \int_{-y_1}^{y_1} \int_{-z_1}^{z_1} 
\sqrt{g}\, d^4 x} \nonumber \\
&=& \lim_{\stackrel{t_1\rightarrow\infty}{t_0\rightarrow 0}}
3 \frac{\int_{t_0}^{t_1} a^2 \left(\ddot{a} -\frac{\dot{a}^2}{a}\right)\, dt}
{\int_{t_0}^{t_1} a^3 \, dt}.
\end{eqnarray}
For both realistic cases of a dust-filled universe ($a(t) = Ct^{2/3}$) and
a radiation-filled universe ($a(t) = Ct^{1/2}$),
$\langle \dot{\theta}\rangle$ vanishes. Indeed, one has
\begin{equation}
\langle \dot{\theta}\rangle_{\rm rad} = 
\lim_{\stackrel{t_1\rightarrow\infty}{t_0\rightarrow 0}}
-\frac{3}{2} 
\frac{\int_{t_0}^{t_1} t^{-1/2}\, dt}
{\int_{t_0}^{t_1} t^{3/2}\, dt} = 0,
\end{equation}
and
\begin{equation}
\langle \dot{\theta}\rangle_{\rm dust} =
\lim_{\stackrel{t_1\rightarrow\infty}{t_0\rightarrow 0}}
- {2} \frac{\int_{t_0}^{t_1} \, dt}
{\int_{t_0}^{t_1} t^2\, dt} = 0.
\end{equation}
As we can see, for both cases, the spacetime averages of the quantities
in the R.H.S. of (\ref{eq}), including the energy density average
$\langle\left(T_{ab} -\frac{1}{2} g_{ab}T \right)\xi^a \xi^b \rangle$,
vanish.

For realistic closed Robertson-Walker universes ($\kappa =1$), we do have
finite spacetime averages. Nevertheless, as it was noted above, one cannot
use the vanishing of the stress energy invariants to decide between 
singularity-free and singular descriptions of the universe.

This work was supported by FAPESP and CNPq.

\begin{flushleft}
Alberto Saa\\
Departamento de Matem\'atica Aplicada, \\
UNICAMP -- C.P. 6065, \\
13081-970 Campinas, SP, Brazil
\end{flushleft}

\end{document}